\documentstyle[aps,prl,floats,epsf,twocolumn]{revtex}

\begin{document}

\draft

\title{Observation of Quantized Hall Drag in a Strongly Correlated Bilayer Electron System}

\author{M.~Kellogg$^1$, I.~B.~Spielman$^1$, J.~P. Eisenstein$^1$, 
L.~N. Pfeiffer$^2$, and K. W. West$^2$}
\address{$^1$California Institute of Technology, Pasadena CA 91125 \\
         $^2$Bell Laboratories, Lucent Technologies, Murray Hill, NJ 07974}

\maketitle

\begin{abstract}
The frictional drag between parallel two-dimensional electron systems has been 
measured in a regime of strong interlayer correlations.  When the bilayer system 
enters the excitonic quantized Hall state at total Landau level filling factor 
$\nu_T=1$ the longitudinal component of the drag vanishes but a strong Hall 
component develops.  The Hall drag resistance is observed to be accurately 
quantized at $h/e^2$.
\end{abstract}


The repulsive interactions between electrons in double layer two-dimensional 
electron systems (2DES) can lead to the condensation, at high magnetic field, of 
a remarkable quantum fluid\cite{perspectives}. The correlations present in this 
fluid include the binding of electrons in one layer to holes in the other.   The 
holes, which in this case are in the conduction band of the host semiconductor 
crystal, exist when the individual 2DES partially fill the discrete Landau 
energy levels produced by a magnetic field applied perpendicular to the 2D 
planes.   From this perspective, the system may be viewed as a Bose condensate 
of interlayer excitons\cite{macd,macd2}. This collective state exhibits the 
quantized Hall effect\cite{eisenstein,lay,murphy} (with Hall 
resistance $R_{xy}=h/e^2$) and has recently been found to display Josephson-like 
interlayer tunneling characteristics\cite{ian1}.  Here we report the observation 
of yet another intriguing property of this system: the exact quantization of the 
frictional drag which one 2DES exerts upon the other.  This frictional drag, 
whose signature is a voltage build-up in one layer in response to a current 
flowing in the other, depends directly on the interlayer correlations present in 
the system.

The excitonic condensate point of view is not unique. The strongly correlated bilayer
system may be described in several mathematically equivalent ways, including as an 
easy-plane ferromagnet or a condensate of composite bosons. In all cases, however, 
the essential physical attribute of the system is interlayer phase 
coherence\cite{yang,moon,perspectives}:  Each electron in the ground state is in 
a specific quantum state which is a linear combination of the individual layer 
eigenstates, $|\!\uparrow \rangle$ and $|\!\downarrow \rangle$.  There is complete 
uncertainty as to which layer any electron (or hole) is in.  If tunneling between 
the layers is strong, this phase coherence is easy to understand: individual electrons 
have lowest energy when they occupy the symmetric double well state 
$|\!\uparrow \rangle + |\!\downarrow \rangle$.  On the other hand, when tunneling is 
weak (or even absent) Coulomb interactions can spontaneously produce interlayer phase 
coherence, provided that the distance separating the two 2DES is less than a critical 
value and the total number of electrons in both layers equals the number of degenerate 
states in the lowest Landau level. In the weak tunneling limit appropriate here,
interlayer phase coherence implies the possibility of superfluid (i.e. dissipationless) 
flow of the excitonic condensate\cite{yang,moon,wen,ezawa,stern}.  However, unlike 
Cooper pairs in a superconductor, interlayer excitons are charge neutral and thus their 
uniform flow corresponds to equal but opposite electrical currents in the two 2DES 
layers. The data presented here provide indirect evidence for the existence of such 
superfluid counterflows.

The samples used in these experiments are GaAs/AlGaAs heterostructures grown by 
molecular beam epitaxy (MBE).  Two 18nm GaAs quantum wells are separated by a 9.9nm 
$\rm{Al_{0.9}Ga_{0.1}As}$ barrier layer.  This double quantum well (DQW) is 
symmetrically doped via Si layers placed in the $\rm{Al_{0.3}Ga_{0.7}As}$ 
cladding layers outside the DQW.  The as-grown density of each 2DES is 
$N_1$=$N_2$=$\rm{5.3\times10^{10}cm^{-2}}$ and their low temperature mobility is 
about $\rm{7.5\times10^5cm^2/Vs}$.  The densities can be independently  varied 
using metal gate electrodes deposited on the sample top surface and back side, 
but for simplicity we shall discuss only the balanced ($N_1$=$N_2$) case here.  
Standard photolithography was used to pattern a square mesa $250\mu \rm{m}$ on a 
side onto the sample. Ohmic contacts were placed at the ends of arms extending 
outward from this mesa.  A selective depletion scheme\cite{contacts} allows 
these contacts to be connected $in~situ$ to either 2DES separately, to both in 
parallel, or to be disconnected entirely.   At zero magnetic field the 
interlayer tunneling resistance of these samples exceeds $30\rm{M}\Omega$.
Data from two, identically patterned, samples cut from the same parent MBE wafer 
are presented here.   

At high magnetic field $B$ the degeneracy $eB/h$ of the lowest spin-split Landau 
level exceeds the electron density $N_{1,2}$ in either layer.  If, however, the 
{\it total} Landau level filling fraction $\nu_T$=$h(N_1+N_2)/eB$ equals unity, 
then the net bilayer system will display the quantized Hall effect (QHE) if the 
layer separation $d$ is small enough or the tunneling is strong enough.   The 
latter case is relatively uninteresting since the origin of the energy gap which 
engenders the QHE is then merely the single-particle tunnel splitting 
$\Delta_{SAS}$ between the lowest symmetric and antisymmetric combinations of 
individual layer eigenstates.  Since the estimated $\Delta_{SAS}$ in the present 
sample is only $\sim\!\rm{0.1mK}$, far smaller than both the measurement 
temperature ($T\!\sim\!\rm{50mK}$) and the mean Coulomb energy ($E_C\!\sim\! 
\rm{50K}$), this mechanism can be safely ignored.  On the other hand, at 
$\nu_T$=1 the excitonic condensate and its associated QHE can develop even in the 
total absence of tunneling if the layers are close enough together\cite{perspectives}. The 
center-to-center quantum well separation $d$=27.9nm of the 
present sample is too large for this to occur at the as-grown densities of the 
2DES. However, since the physics is governed by the ratio of $d$ to the average 
separation between electrons within each layer, the transition to the excitonic 
phase can be driven by reducing the densities $N_{1,2}$.  At fixed filling 
fraction the mean electron spacing is simply proportional to the magnetic length 
$\ell$=$(\hbar /eB)^{1/2}$ = $(\nu_T /2 \pi N_T)^{1/2}$.  Via gating we are able 
to reduce the key ratio $d/\ell$ at $\nu_T$=1 from about 2.3 down to below 1.6.  
Consistent with earlier observations\cite{ian1}, the $\nu_T$=1 bilayer QHE first 
appears around $d/\ell$=1.83.  By $d/\ell$=1.6 it is well-developed: A deep 
minimum is observed in the longitudinal resistance $R_{xx}$ and a clear plateau 
is evident in the Hall resistance at $R_{xy}=h/e^2$. In this situation the QHE 
is due almost exclusively to electron-electron interactions.  
 
Frictional drag measurements\cite{gramila,sivan,lilly} are performed by driving 
current through one 2DES while monitoring the voltage which appears in the 
other, electrically isolated, 2DES.  The drag voltage is a direct measure of the 
interlayer momentum relaxation rate\cite{gramila,jauho}.  In the present sample, 
with its small layer separation and low electron density, the dominant 
relaxation mechanism at low temperatures is direct electron-electron Coulomb 
scattering.  A careful study, to be reported elsewhere, of the zero magnetic 
field drag in these samples reveals the expected near-quadratic temperature 
dependence. For reference, at $N_1$=$N_2$=$5.3\times10^{10}\rm{cm^{-2}}$ the 
measured drag resistivity is $\rho_D\approx(0.4\Omega/\Box$-$\rm{K^2})$$T^2$ for 
$T<4\rm{K}$.

\begin{figure} [h]
\begin{center}
\epsfxsize=3.3in
\epsffile[105 224 503 571]{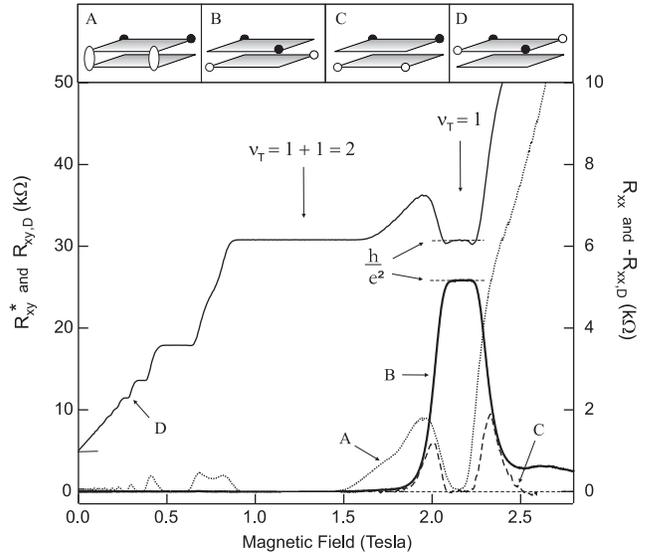}
\end{center}
\caption[figure 1]{Conventional and Coulomb drag resistances of a low 
density double layer 2DES.  Trace A: Conventional longitudinal resistance 
$R_{xx}$ measured with current in both layers.  Trace B: Hall drag resistance 
$R_{xy,D}$. Trace C: Longitudinal drag resistance $R_{xx,D}$; sign reversed for
clarity. Trace D: Hall 
resistance ${R_{xy}}^*$ of single current-carrying layer (displaced vertically 
by 5k$\Omega$ for clarity). Trace B reveals the quantization of Hall drag in the 
$\nu_T$=1 excitonic QHE. Insets schematically illustrate the measurement 
configurations: Current is injected and withdrawn at the open dots; voltage 
differences between the solid dots are recorded.  Traces A,B, and D obtained at 
$T=20$mK; trace C at 50mK.  Layer densities: 
$N_1$=$N_2$=$2.6\times10^{10}\rm{cm^{-2}}$, giving $d/\ell$=1.6 at $\nu_T$=1.
}
\end{figure}

Figure 1 shows the main results of this study. The densities have been reduced 
by symmetric gating to $N_1$=$N_2$=$2.6\times10^{10}\rm{cm^{-2}}$, making 
$d/\ell$=1.6 at $\nu_T$=1.   The four traces shown correspond to the magnetic 
field dependence of various voltage measurements made on the system; these are 
converted to resistances by dividing by the excitation current $I$, typically 
2nA at 5Hz.   The insets to the figure depict the various measurement 
configurations. Trace A shows the conventional longitudinal resistance $R_{xx}$ 
of the sample, measured with the current flowing in parallel through both 
layers.  The deep minimum near $B$=2.15T reflects the strong $\nu_T$=1 bilayer 
QHE present at this density. Although omitted from the figure, a well-developed 
plateau is also observed in the conventional Hall resistance of the sample at 
$R_{xy}=h/e^2$.

Traces B and C illustrate our most important results.  For these data the 
excitation current was driven through only one 2DES while voltages in the non-current-carrying 2DES were recorded.  Trace B represents ``Hall drag", a voltage 
drop which appears transverse to the current flowing in the other layer.   At 
low magnetic field the Hall drag resistance $R_{xy,D}$ is undetectably small but 
around $B$=2T it rises up and forms a flat plateau.  This plateau is centered at 
the location of the $\nu_T$=1 QHE state.  At still higher fields $R_{xy,D}$ falls 
off again to much smaller values.  On the plateau we have found that $R_{xy,D}$ 
equals the quantum of resistance $h/e^2$=25,813$\Omega$ to within about 5 parts in 
$10^4$. We emphasize that the same quantization of Hall drag is observed when the 
roles (drive $vs.$ drag) of two layers are interchanged and that the sign of the 
Hall drag is the same as that of the conventional Hall effect in the current-carrying 
layer.

Along with this plateau in the Hall drag, trace C demonstrates that the longitudinal 
drag resistance $R_{xx,D}$ (i.e. the drag voltage drop which is parallel to the 
current in the drive layer) simultaneously exhibits a deep minimum.  Note, however, 
that the longitudinal drag voltage is opposite in sign to the longitudinal resistive 
voltage drop in the current-carrying layer.  This sign difference (which has been 
removed for clarity from Fig. 1) is commonplace is drag studies\cite{gramila,lilly}
on weakly correlated
bilayer electron systems where it merely refects the force balance resulting from the 
constraint that no current flow in the drag layer. In any case, it is apparent from 
Fig. 1 that the two components of Coulomb drag display the $\nu_T$=1 quantized Hall 
effect just as conventional resistivity measurements do, in spite of the fact that the 
drag voltages exist in the layer in which there is no current. 

Finally, for trace D the current again flows through only one layer but now the 
Hall voltage across that same layer is recorded.  At low magnetic fields this 
Hall resistance, denoted by ${R_{xy}}^*$, reflects \it single-layer \rm physics:  
The slope of the initial linear rise of ${R_{xy}}^*$ with field is determined by 
the density of the current-carrying layer ($N_{1,2}$=$N_T/2$) and the 
subsequent QHE plateaus at intermediate fields correspond to integer values of 
the individual filling factors $\nu_{1,2}$.   The last such single-layer QHE 
plateau, at ${R_{xy}}^*$=$h/e^2$, is centered at $B$=1.1T and corresponds to 
$\nu_1$=$\nu_2$=1, i.e. $\nu_T$=2.  At still higher fields ${R_{xy}}^*$ 
begins to deviate from $h/e^2$ but then remarkably returns to form a {\it second 
plateau} at $h/e^2$ around $B$=2.15T, exactly where the Hall drag plateau 
exists and the bilayer system is in the $\nu_T$=1 QHE state.

The assumption that no current flows in the layer in which drag voltages are 
measured is always a key issue in drag experiments. It requires particularly 
careful scrutiny at $\nu_T$=1 since a huge increase in the interlayer tunneling 
conductance has been observed\cite{ian1} to occur when $d/\ell$ is reduced below 
$\sim$1.83 and the excitonic condensate develops.  Several facts, however, leave
us confident that tunneling is not a serious problem.  First, the tunneling enhancement
is sharply resonant around zero interlayer voltage.  At low temperatures the width of
the tunnel resonance in the present samples is less than $10\mu{\rm V}$\cite{ian2}. 
In contrast, we find the Hall drag plateau unaffected by intentionally imposed
interlayer voltages of up to $\pm500\mu{\rm V}$. Second, a small additional magnetic 
field ($B_\parallel$$\sim0.7$T) applied parallel to the 2D layers has been 
demonstrated\cite{ian2} to suppress the $\nu_T=1$ tunneling conductance by more 
than an order of magnitude.  We find that the same in-plane field has no effect 
on the quantized Hall drag plateau.  Finally, direct tunneling experiments on the present
samples have shown that the {\it maximum} tunnel current that can flow between the layers 
at $\nu_T=1$ is around 10pA, independent of interlayer voltage up to several mV.  Since 
our drag measurements are performed with excitation currents of $\sim 1{\rm nA}$, a 
reasonable worst-case estimate of the maximum current flowing in the ``wrong" layer 
is 1\% of the total. 

\begin{figure}
\begin{center}
\epsfxsize=3.3in
\epsffile[112 232 481 531]{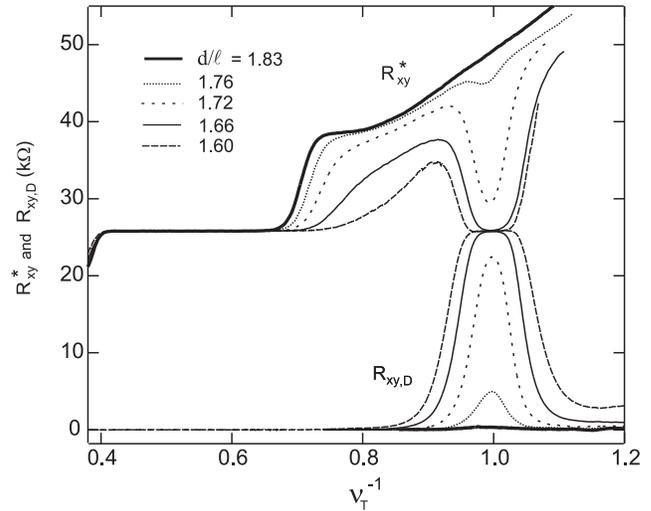}
\end{center}
\caption[figure 2]{Collapse of $\nu_T$=1 Hall drag quantization and second 
$h/e^2$ plateau in ${R_{xy}}^*$ at large $d/\ell$.  Layer densities 
$N_1$=$N_2$=2.6, 2.8, 3.0, 3.2, and $3.4\times10^{10}\rm{cm^{-2}}$, giving 
$d/\ell$=1.6, 1.66, 1.72, 1.76 and 1.83 respectively, at $\nu_T=1$. Measurement 
temperature $T$=30mK.
}
\end{figure}
The data in Fig. 1 demonstrate that the same Hall voltage appears across both 
layers at $\nu_T$=1, in spite of the fact that current flows only in one of 
them.  This voltage is precisely the same as that which appears across both 
layers when the current is driven in parallel through both layers.  Thus, the 
same voltages appear across both layers irrespective of how the total current 
$I$ is divided between them.  This remarkable fact is a direct manifestation of 
interlayer phase coherence.

Figure 2 shows that the phenomena of quantized Hall drag and the anomalous 
second $h/e^2$ plateau in ${R_{xy}}^*$ both disappear when the effective layer 
separation $d/\ell$ is increased beyond about 1.83.  To facilitate their 
comparison, the data in Fig. 2 are plotted versus inverse total filling factor 
$\nu^{-1}_T$, 
not magnetic field.  Not surprisingly, at large $d/\ell$ very little Hall drag 
is present and the Hall resistance ${R_{xy}}^*$ of the current-carrying layer 
remains close to the classical Hall line in the field range around $\nu_T$=1.  
Although not shown in the figure, the minimum in the longitudinal drag 
$R_{xx,D}$ at $\nu_T$=1 is also absent at large $d/\ell$.  To within 
experimental uncertainty, the collapse of quantized Hall drag occurs 
simultaneously with the vanishing of the conventional QHE and the system's
Josephson-like tunneling characteristics. 

Figure 3 displays the temperature dependence of these phenomena, again at 
$d/\ell$=1.6. Three data sets are shown: The conventional longitudinal 
resistance $R_{xx}$ (measured with current flowing in parallel through both 
layers), the longitudinal drag resistance $R_{xx,D}$, and the deviation $\Delta 
R_{xy,D}$ of the Hall drag from its quantized value of $h/e^2$.   As the figure 
shows, each of these quantities is approximately thermally activated [i.e. is 
proportional to exp(-$E_A/T$)] at low temperatures.  As the near parallel slopes
suggest, the activation energies are all comparable: $E_A$$\approx 0.4{\rm K}$.

\begin{figure}
\begin{center}
\epsfxsize=3.3in
\epsffile[0 0 324 263]{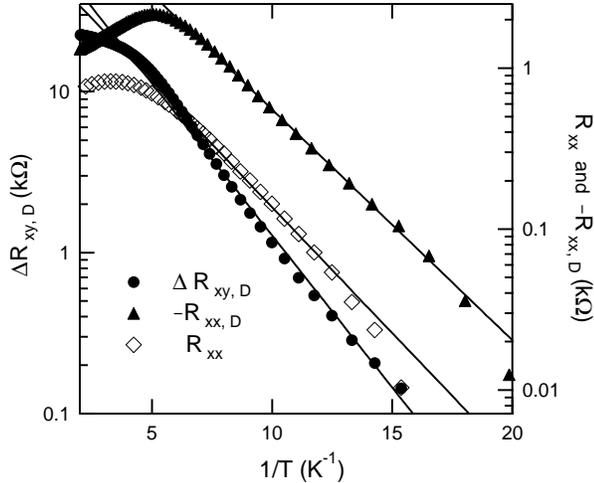}
\end{center}
\caption[figure 3]{Temperature dependences of conventional longitudinal 
resistance $R_{xx}$, longitudinal drag resistance $R_{xx,D}$, and deviation 
$\Delta R_{xy,D}$ of the Hall drag from $h/e^2$ at $\nu_T$=1 and $d/\ell$=1.6.
The sign of $R_{xx,D}$ has been reversed for clarity. The lines are guides 
to the eye. 
}
\end{figure}

The existence of quantized Hall drag is remarkable.  At the simplest level, one
expects it should not exist.  For two uncorrelated 2D layers the usual argument is that
since no current flows in the drag layer, there can be no Lorentz force on its carriers.
Without the Lorentz force, there ought not be any voltage build up transverse to
the current in the drive layer.  This argument, however, is fallacious, even without
interlayer correlations.  Hu\cite{hu}, and later von Oppen {\it et al.}\cite{vonOppen},
showed that Hall drag voltages can exist provided there is an energy (or density) dependence
to the carrier momentum relaxation rate.  More relevant here however, are the several
theoretical predictions of large and {\it quantized} Hall drag voltages that result from 
strong interlayer correlations\cite{moon,renn,duan,yang2,yang3,kim}. For example, 
Yang\cite{yang2} has recently shown that quantized Hall drag at $\nu_T$=1 follows from 
the assumption of the specific many-body ground state wavefunction\cite{halperin} (the 
so-called 111-state) generally believed to capture the essential physics of spontaneous 
interlayer phase coherence and exciton condensation. 

Interlayer phase coherence implies that electrons are spread equally between both 
layers. A non-equilibrium current injected into one layer thus divides equally into 
both layers and the resulting Hall voltages in the two layers are the same.  On the 
other hand, this current division obviously violates the basic boundary conditions 
of a drag measurement.  According to theory\cite{vignale}, the resolution of this 
paradox lies in the superfluid properties of the excitonic condensate itself.  In 
addition to the transport current flowing equally through both layers, a superflow of 
excitons develops.  Since such a superflow corresponds to counterflowing electrical 
currents in each layer, it produces no Hall field and allows for the net current in 
one layer to be zero while in the other layer it is finite. Only if the net currents 
in the two layers are equal is there no such superflow. From this perspective our 
experimental results offer the first, albeit indirect, evidence for excitonic 
superfluidity at $\nu_T$=1. 

It is a pleasure to acknowledge valuable discussions with S. Das Sarma, S.M. 
Girvin, A.H. MacDonald, A. Stern, and K. Yang.  This work was supported 
in part by grants from the NSF and the DOE.  One of us (I.B.S.) acknowledges the 
support of the Department of Defense.


\begin{references}

\bibitem{perspectives} For a review, see the chapter by S.M. Girvin and A.H. 
MacDonald in {\it Perspectives in Quantum Hall Effects}, edited by S. Das Sarma 
and A. Pinczuk, (John Wiley, New York, 1997).

\bibitem{macd}A.H. MacDonald and E.H. Rezayi, Phys. Rev. B{\bf 42}, 3224 (1990).

\bibitem{macd2}A.H.MacDonald, Physica B {\bf 298}, 129 (2001). 

\bibitem{eisenstein}J.P. Eisenstein, G.S. Boebinger, L.N. Pfeiffer, K.W. West, 
and S. He, Phys. Rev. Lett. {\bf 68}, 1383 (1992).

\bibitem{lay}T.S. Lay, {\it et al.} Phys. Rev. B{\bf 50}, 17725 (1994).

\bibitem{murphy}S.Q. Murphy, J.P. Eisenstein, G.S. Boebinger, L.N. Pfeiffer, 
and K.W. West, Phys. Rev. Lett. {\bf 72}, 728 (1994).

\bibitem{ian1}I.B. Spielman, J.P. Eisenstein, L.N. Pfeiffer, and K.W. West, 
Phys. Rev. Lett. {\bf 84}, 5808 (2000).

\bibitem{yang}Kun Yang, {\it et al.}, Phys. Rev. Lett. {\bf 72}, 732 (1994).

\bibitem{moon}K. Moon, {\it et al.} Phys. Rev. B {\bf 51}, 5138 (1995).

\bibitem{wen}X.G. Wen and A. Zee, Phys. Rev. Lett. {\bf 69}, 1811 (1992).

\bibitem{ezawa}Z.F. Ezawa and A. Iwazaki, Phys. Rev. B{\bf 47}, 7295 (1993).

\bibitem{stern}A. Stern, S. Das Sarma, M.P.A. Fisher, and S.M. Girvin, Phys. Rev. Lett.
{\bf 84}, 139 (2000).

\bibitem{contacts}J.P.Eisenstein, L.N. Pfeiffer, and K.W. West, Appl. Phys. Lett.
{\bf 57}, 2324 (1990).

\bibitem{gramila}T.J. Gramila, J.P. Eisenstein, A.H. MacDonald, L.N. Pfeiffer, and
K.W. West, Phys. Rev. Lett. {\bf 66}, 1216 (1991).

\bibitem{sivan}U. Sivan, P.M. Solomon, and H. Shtrikman, Phys. Rev. Lett. {\bf 68}, 1196 (1992).

\bibitem{lilly}M.P. Lilly, J.P. Eisenstein, L.N. Pfeiffer, and K.W. West, Phys. Rev. Lett. 
{\bf 80}, 1714 (1998).

\bibitem{jauho}A.P. Jauho and H. Smith, Phys. Rev. B{\bf 47}, 4420 (1993).

\bibitem{ian2}I.B. Spielman, J.P.Eisenstein, L.N. Pfeiffer, and K.W. West, Phys. Rev. Lett.
{\bf 87}, 036803-1 (2001).

\bibitem{hu}B.Y.K. Hu, Physica Scripta {\bf T69}, 170 (1997).

\bibitem{vonOppen}F. von Oppen, S. H. Simon, and A. Stern, Phys. Rev. Lett. {\bf 87}, 106803 (2001).

\bibitem{renn}S.R. Renn, Phys. Rev. Lett. {\bf68}, 658 (1992).

\bibitem{duan}J-M. Duan, Europhys. Lett. {\bf 29}, 489 (1995).

\bibitem{yang2}Kun Yang, Phys. Rev. B{\bf 58}, R4246 (1998).

\bibitem{yang3}Kun Yang and A.H. MacDonald, Phys. Rev. B{\bf63}, 073301 (2001).

\bibitem{kim}Y.B. Kim, C. Nayak, E. Demler, N. Read, and S. Das Sarma, Phys. Rev. B{\bf 63}, 205315 (2001).

\bibitem{halperin}B.I. Halperin, Helv. Phys. Acta {\bf 56} 75 (1983).

\bibitem{vignale}G. Vignale and A.H. MacDonald, Phys. Rev. Lett. {\bf76}, 2786 (1996).

\end{references}
\end{document}